# Amorphous state of $sp^2$ solid carbon

E. F. Sheka *[1], Ye. A. Golubev [2], N. A. Popova [1]

[1] *Peoples' Friendship University (RUDN University) of Russia, Miklukho-Maklay, 6, 117198 Moscow, Russia*
[2] *Yushkin's Institute of Geology, Komi Science Center, Ural Branch of RAS, Pervomayskaya, 54, 167982, Syktyvkar, Russia*

*Corresponding authors: e-mail: sheka@icp.ac.ru (Elena Sheka)

**ABSTRACT**. Two-mode valence electron configuration of carbon atoms lays the foundation of the unique two-mode amorphous state of the monoatomic carbon solid. From the fundamentals of solid-state physics, $sp^3$ and $sp^2$ amorphous carbons are two different amorphous species characterized by conceptually different short-range orders, namely, groups of tetrahedrally bonded $sp^3$ configured atoms and nanoscale size-restricted $sp^2$ graphene domains framed with heteroatom necklaces. Molecular character of $sp^2$ amorphics is coherent with the reaction mechanism of the solid amorphicity in due course of the enforced fragmentation.

## I. NEW GRAPHENE STORY OF OLD AMORPHOUS CARBON

Recent purposeful studies [1–4] have completed the gradually nascent change in the view of $sp^2$ amorphous carbon as a well-known, familiar physicochemical subject, the beginning of which was laid back in 1941 [5], transferring it to the rank of *high-tech* material of modern nanotechnology. We are talking about material known to humankind since the first decomposed bonfire, which left behind black soot and charcoal. Today, this applies to billions (anthracite coal) - millions (shungite carbon) - thousands (anthraxolite and blackcarbon coatings, which are present everywhere and accompany almost all minerals in nature) tons of only discovered natural deposits and hundreds of millions of tons of synthetic black carbon produced industrially. All this black carbon wealth is amorphous carbon, or a monoatomic solid without long-range order, the atoms of which form $sp^2$ configured valence bonds with each other. Because of scrupulous studies [1-4], it was finally established that the black carbon richness described above has a unique common basis, namely, nano-micro-scale molecular compositions of hexagonal honeycomb structures of carbon atoms (graphene domains) framed by various heteroatom necklaces formed by oxygen, hydrogen, nitrogen, sulfur, halogen and other atoms. These framed graphene molecules are basic structural elements (BSUs), varying in size and shape, as well as differing in the chemical composition of the necklaces depending on the history or method of origin and/or production of the black carbon described above. The discovered and experimentally confirmed, graphene nature of this black gold leads to a revolutionary revision of the theory, modeling and interpretation of the experiment related to this class of solids, which concerns:

*Experiment:* multilevel fractal structure, accompanied by pronounced porosity; BSU graphene domains; heteroatom framing necklaces - the nature of their occurrences and functions;

*Theory:* class of amorphous solids - amorphics with a molecular structure; BSU of variable size as a carrier of size-sensitive effects; molecular-crystalline dualism with respect to the spectra of electronic and vibrational states; molecular and solid-state quasiparticle approximations as research methods;

*Modeling:* nanoscale framed graphene molecules (real BSUs) as objects of modeling; the radical nature of BSU; size-kinetic nature of stabilization of BSU molecular radicals; manifestation of the radical nature of BSU;

*Practical applications* based on spin chemistry, carbocatalysis, spin tribology, spin geochemistry, and spin medicine of $sp^2$ graphenous amorphous carbons.

All the issues present the modern content of the current material science of the $sp^2$ amorphous carbons and many successful results can be found in literature related to each of them to construct the building of this peculiar material. However, there are still some unanswered fundamental questions which prevent from the construction and those are the following: (i) what is amorphous state of the solid carbon from the standpoint of general solid-state concepts; and (ii) what is the place and role of $sp^2$ amorphous carbons in the total picture of the carbon amorphicity. The current paper is aimed at answering these questions.

## II. TODAY'S PRESENTATION OF AMORPHOUS CARBON

Amorphous carbons (ACs) are widely common and present a big allotropic class of bodies, both natural and synthetic. Natural amorphics are products of the activity of the Nature's laboratory during geological billion-million-year time. Exhausted geological examinations allowed suggesting a few classification schemes of carbon species (see [6-11] but a few), one of which, to whose updated form geologists are returning today, slightly changed with respect to original [6], is schematically shown in Fig. 1a. The scheme presents a continuous evolution of pristine carbonaceous masses into ordered crystalline graphite thus exhibiting the main stream of carbon life in the Nature. The evolution is presented as increasing carbonization rank of intermediate products. As seen in the figure, a general picture can be split into two gloves, the left of which starts with plants and sediments of different kinds and proceeds through sapropels to brown coals and later to convenient coals and anthracite. The endpoint on the way is graphite. As for the right glove, it covers carbonization of pristine gas and distillate oil and proceeds through petroleum and naphthoids to asphalts and then to kerites, anthraxolites, and shungites. As in the previous case, graphite is the endpoint. Certainly, the division is not exactly rigid, due to which a mixture of the two fluxes, particularly, at early stages of carbonization, actually occurs. This scheme is related to $sp^2$ amorphous carbon that actually dominates in the Nature. Natural $sp^3$ amorphous carbon is not so largely distributed, due to which diamond-like natural ACs have become top issues of the carbon mineralogy [12, 13] only recently.

The family of synthetic amorphous carbons is quite large, covering species different by not only the carbonization rank, but also a mixture of $sp^2$ and $sp^3$ components. Analogously to natural species, synthetic amorphics were classified as well [14, 15] and the relevant classification scheme is presented in Fig. 1b. Previously ternary, in the current study, it is completed to a rhombic one to take into account oxygen as one other important ingredient, which is caused by large development of modern techniques of the ACs production. Comparing schemes presented in Figs. 1a and 1b, it becomes evident that those are related to fully different communities of substances. If natural amorphics belong to $sp^2$ carbon family and are rank-characterized with respect to the stage of their metamorphism and carbonization, synthetic amorphics are mainly characterized by $sp^3$-configured solid carbon. $sp^2$ Group of the synthetic species takes only small place, marked by oval, in the total family, characterized by the highest carbonization. Because of a small amount of $sp^3$ solid amorphous carbon in nature, special technologies to produce tα-C, tα-CH (t means tetragonal), and sputtered $sp^2$ & $sp^3$ mixed α-C:H products were developed. It is necessary to complement this part of carbon solids by graphene oxide which, in ideal case, corresponds to $sp^3$ configured carbon only and to about 30 wt % of oxygen and which cannot be presented on the considered planar classification scheme. As for $sp^2$ synthetic amorphics, for a long time they were presented by multi-tonnage industrial production of glassy (covering graphitic, black, activated and other highly carbonized) carbon [16]. However, the graphene era called to life a new high-tech material – technical graphenes [17] which are the final product of either oxidation-reduction [18] or oxidation-thermally-shocked exfoliation [19] of nanosize graphite. Two new members of this community are on the way – that are laser-induced graphene (LIG) manufactured by multiple lasing on cloth, paper, and food [20] and extreme quality flash graphene (FG) [21], closest in ordering to graphite.

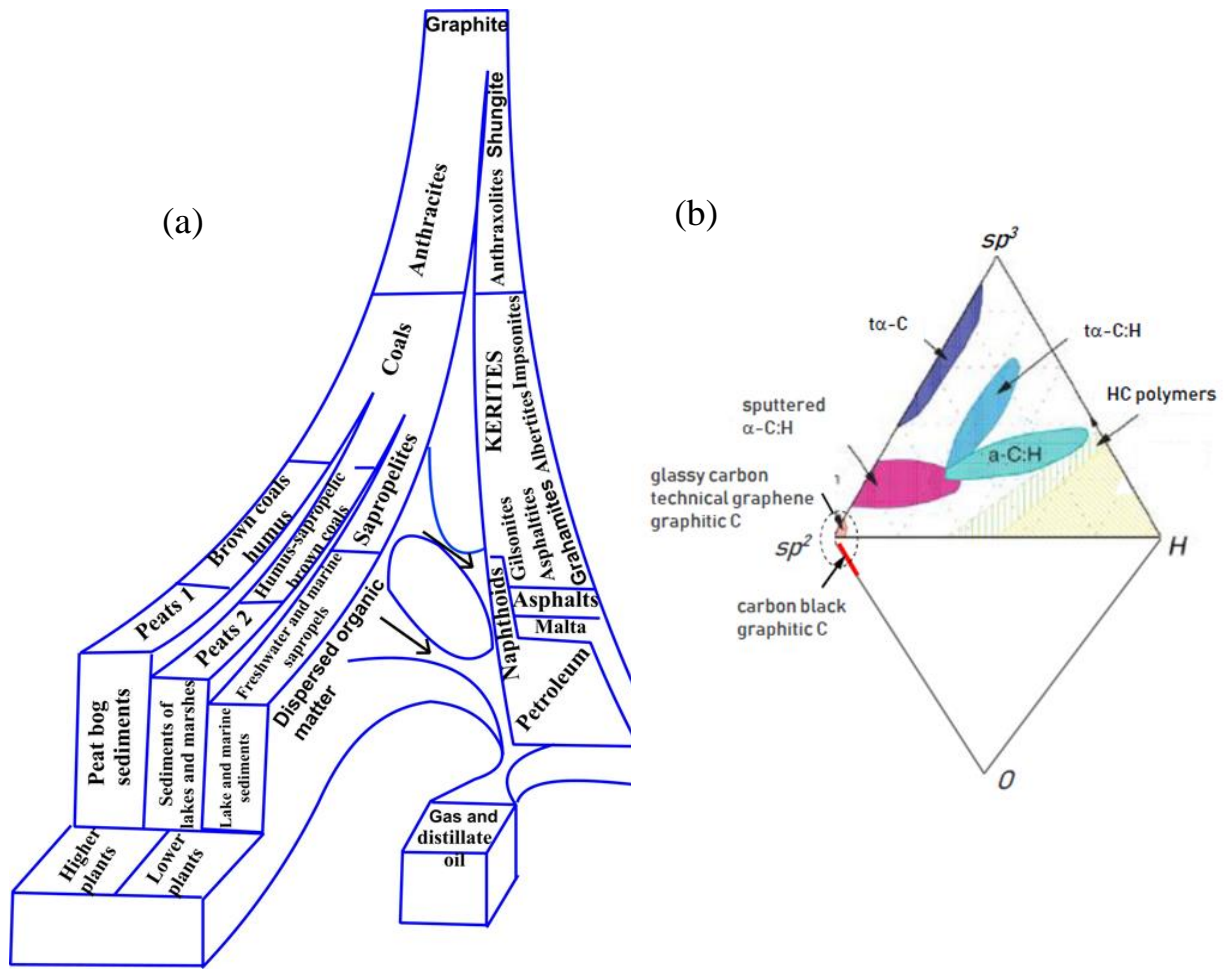

**Figure 1.** a. The path of carbon life in the Nature: Amorphous carbons based on the Uspenskiy's classification [6]. b. Rhombic phase diagram of synthetic amorphous carbon-hydrogen-oxygen system.

Therefore, $sp^2$ ACs constitute the main part of amorphous carbon. Despite they have been the object of study and practical use for hundreds of years, until now they have not been considered from the general concept of the solid-state physics. The amorphicity of solids was widely studied and the main concepts are accumulated in monograph [22]. The first conceptual issue concerns a considerable degree of amorphous solid ordering that is subdivided into short-range (local) and medium range ones, the boundary between which passes around a few nm. The second issue is related to the direct interconnection of the solids properties and their local ordering due to which establishing a local structure has always been the main goal of studying. The peak development of this solid state physics was reached in the third quarter of the previous century. Historically, the most attention has been given to monoatomic Si and Ge, which, it would seem, is quite conducive to our study, because carbon, silicon, and germanium form a common tetrel family so that a similar behavior could be expected to all the members. As was found, tetrahedrally bonded $sp^3$ configured atoms form the short-range order of Si and Ge amorphous solids. However, in the case of carbon, as seen in Fig. 1b, similar amorphous compositions of carbon are concentrated near $sp^3$ corner and are related to t$\alpha$-C phase. The main part of the pure C-amorphics is presented by $sp^2$-configured solids, which do not exist in the case of Si and Ge. We believe that this feature is on line with other exclusivenesses related to the tetrel family, such as the absence of both aromatic families of Si and Ge chemical compounds including fullerenic species, on the one hand, and freestanding honeycomb monolayer silicene and germanene similar to graphene, on the other. The feature is caused by an extreme radicalization of the elongated double covalent bonds X=X (X= Si, Ge) [23, 24], which makes the bodies chemically unstable under ambient conditions. Therefore, from the standpoint of general concepts of the solids amorphicity, monoatomic solid carbon has the unique ability to form amorphous (as well as crystalline) states of two types, characterized by fundamentally different short-range orders presented by either tetrahedral $sp^3$ groups of bonded atoms or an $sp^2$ honeycomb network of benzenoid units, thus differentiating $sp^3$ and $sp^2$ amorphous carbon. The short-range order of the first group amorphics is rather simple and similar to that one widely studied for Si and Ge [16]. In contrast, the case of $sp^2$ amorphics is quite peculiar and must be considered in details.

Recent comparative studies [1-4] together with a large pool of individual data have shown that these amorphics form a particular class of solids. As turned out, the general architecture of both natural and synthetic species is common and can be characterized as multilevel fractal structure [25, 26], albeit differing in details at each level. The amorphics's structure of the first level is well similar in all the cases and is presented by *basic structure units* (BSUs). As mentioned above, the units are framed graphene molecules and/or graphene domains. The framing plays a decisive role, ensuring the formation and stability of short-range order, on the one hand, and preventing crystallization, on the other [27, 28], thus allowing attributing the origination of $sp^2$ ACs to the *reaction amorphization* [22]. The second level structure is provided with nano-thick BSUs stacks, which confidently recorded by X-ray and neutron diffraction structural studies of $sp^2$ ACs of all types [2]. The third-level structure of the amorphics reliably follows from the porous structure evidently observed experimentally [26, 29. It is constructed from the BSUs stacks but the final compositions depends on the stacks lateral dimension. When the latter is at the first nanometer level, the composition presents globules of ~10 nm in size, which corresponds to pores, size of which is first nanometers as well [30]. Further aggregation of globules leads to the formation of micro-nanosize agglomerates with pores of tens nm [26]. Such a structure is typical to natural ACs such as shungite carbon, anthraxolite, anthracite as well as black carbon coating of diamonds [31], mixed carbon-silica spherical 'sweets' [32], black carbon in meteorites [33] and metamorphosed polymictic sandstones [34] and none of the exclusions has been known so far. Figure 2 presents schematically the evolution of amorphic structure of this kind from a single BSU to macroscopic powder. A schematic structure of a globule, shown in the figure, is designed on the basis of data related to shungite carbon. Molecule $C_{66}O_4H_6$ constitutes one of possible models of the specie BSUs based on the graphene domain $C_{66}$. The model is commensurate by size with real BSU and its heteroatom necklace is composed basing on chemical data of the species established in [2]. Combined into four-, five-, and six-layer stacks in relation with the structural data [1-3], the molecules create a visible picture of the nanoscale globules, further agglomeration of which provides the final nanostructured view of the species presented by 3D AFM image in the figure. In contrast to natural bodies, synthetic amorphics are characterized by a large dispersion of BSUs size from units to tens and/over first hundreds of nanometers. At the low-limit end of the dispersion, the amorphic structure is similar to that of natural species described above. At the high-limit end, the BSUs size does not prevent from BSUs packing in nanosize-thick stacks while the latter laterally extended are further packed in a paper-like structure exemplified for technical graphenes in Fig. 3.

The above concept concerning $sp^2$ ACs structure is based on planar BSUs. The latter are indeed characteristic of the real structure and are numerous. However, sometimes the images of natural ACs also contain bent fragments, and some of them form smoothly bent ribbons up to first tens nanometers in size. Since interplanar spacing $d$ for these arched structural fragments is usually 0.34–0.38 nm, the bending cannot be connected with a chemical modification of BSU within its basal plane that might curve the graphene domains. It is reasonable to relate the bending of BSUs to the copresence of various mineral inclusions in deposites, which usually accompany ACs formation, thus, for example, from silica micro- and nanoparticles to metal nanoparticles in the case of shungite carbon [17].

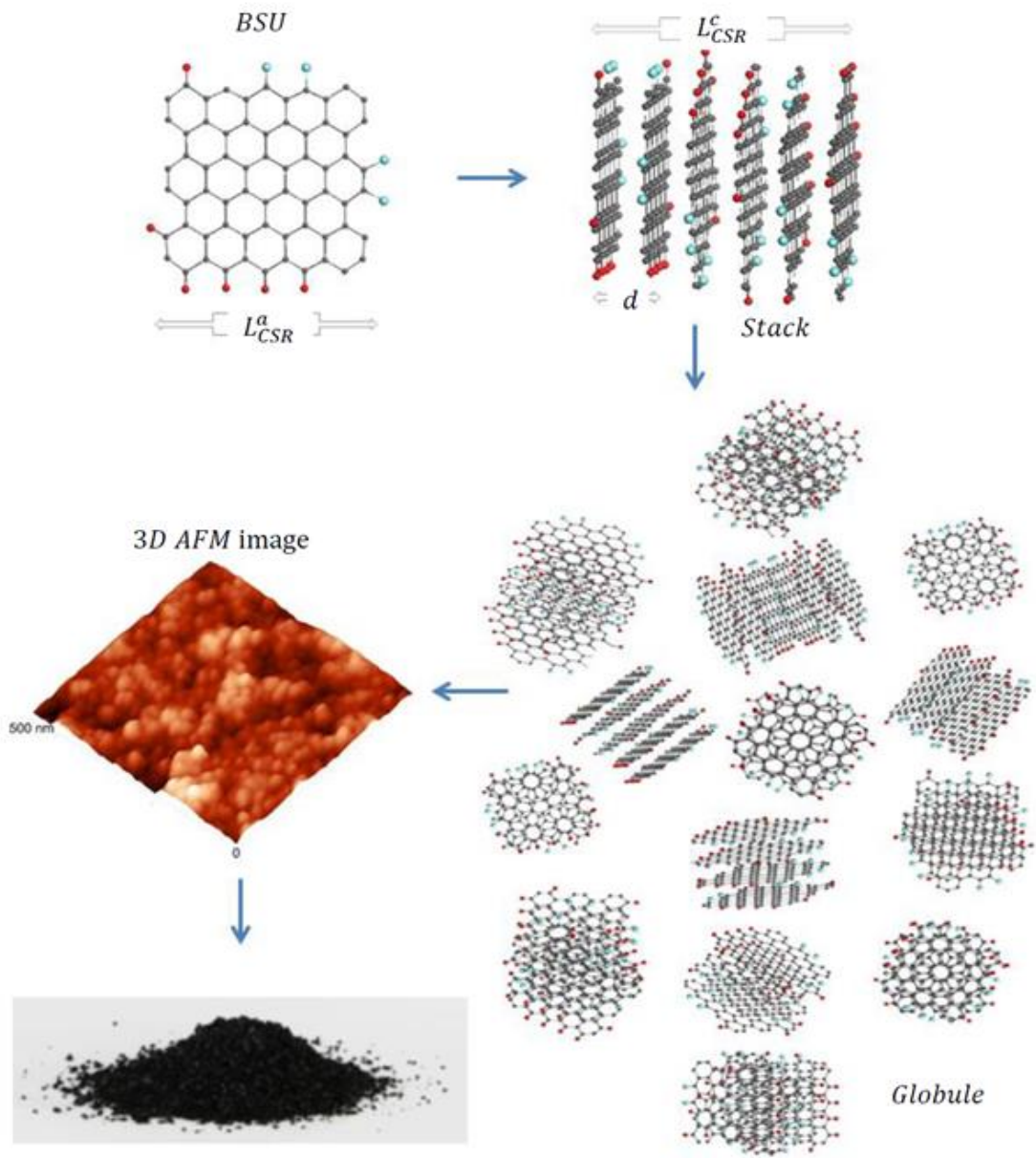

**Figure 2.** Schematic transformation from a single BSU molecule (one of the BSU models $C_{66}O_4H_6$ for shungite carbon [3]) to powdered solid amorphous carbon via BSU stacks and globule(s). The stacks consist of a number of BSU layers from 4 to 7, differently oriented to each other. Planar view on a model globule composed of different stacks, with total linear dimensions of ~ 6 nm. Dark gray, red and blue balls depict carbon, hydrogen, and oxygen atoms, respectively. $L^a_{CSR}$ and $L^c_{CSR}$ are linear size of BSU and BSU stack thickness of 2.0 nm and 2.1 nm, respectively; $d$ is interlayer distance of 0.348 nm [2]. 3D AFM image of globular structure of shungite carbon powder obtained with NTEGRA Prima, NT-MDT. The size distribution of the observed nanoscale motives is peaked at 25 nm.

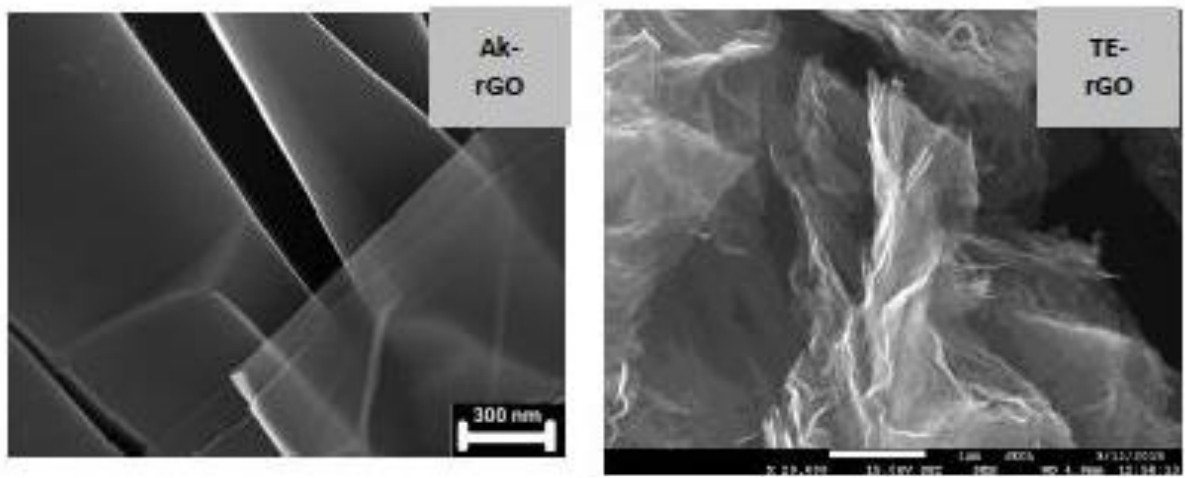

**Figure 3.** SEM images of technical graphene Ak-rGO [35] and TE-rGO [36]. Adabted from Ref. [3].

**III. ENFORCED FRAGMENTATION AS THE AMORPHICITY TYPE OF *sp²* SOLID CARBON**

Continuing the discussion of the general view of the *sp²* carbon amorphicity, we should dwell on one more distinctive feature. Evidently, the molecular nature of BSUs makes it possible to reliably attribute *sp²* ACs to amorphics with a molecular structure. However, in contrast to typical representatives of this class of solids, which are based on molecules of a stable standard structure [38, 39], BSUs of *sp²* ACs are not standard neither structurally nor atomically, so that their structural and chemical-content data, such as obtained in [1-3], represent only statistically averaged quantities. For example, the positions of hydrogen and oxygen atoms in the BUS heteroatom necklace, shown in Fig. 2, can markedly vary. The same concerns the molecule shape and exact size. Besides, in the case of ACs, the transition from disordered amorphous structure to ordered one occurs quite differently. In ordinary amorphics, the ordering proceeds as a gradual accommodation of the positions of constituent molecules with respect to each other via a sequence of mesomorphic transformation thus aligning translational periodicity (see details in [39]). In contrast, the crystallization, better to say, graphitization of *sp²* ACs occurs due to increasing the BSU size, as is shown schematically in Fig. 4. To trace processing, we have fixed the atomic C:O:H relative content by the data related to shungite carbon [2] and have simulated the intermediate transition by growing the graphene domain. Certainly, the size-increasing of individual BSUs is accompanied with bridging of both BSUs themselves and stacks of them. For many years, this bridging was considered as a main process of sequential ordering of the primary block-mosaic structure of natural ACs [40-43]. Obviously, the priority of one of the two processes depends on the environment. Nevertheless, not depending on whichever process dominates in practice, we should come to a conclusion that *sp²* amorphous solid state can not be attributed to any of the known types of disorder characteristic for monoatomic solids [22]. What is observed empirically in the case of ACs, in terms of solid-state physics can be attributed to *enforced fragmentation* of graphite and/or graphene crystals. Obviously, the fragmented solid can be obtained from both the top and bottom. In the first case, it concerns the disintegration of pristine graphite and/or graphene crystal while in the second case it concerns stopping the graphitization of pristine graphene lamellas.

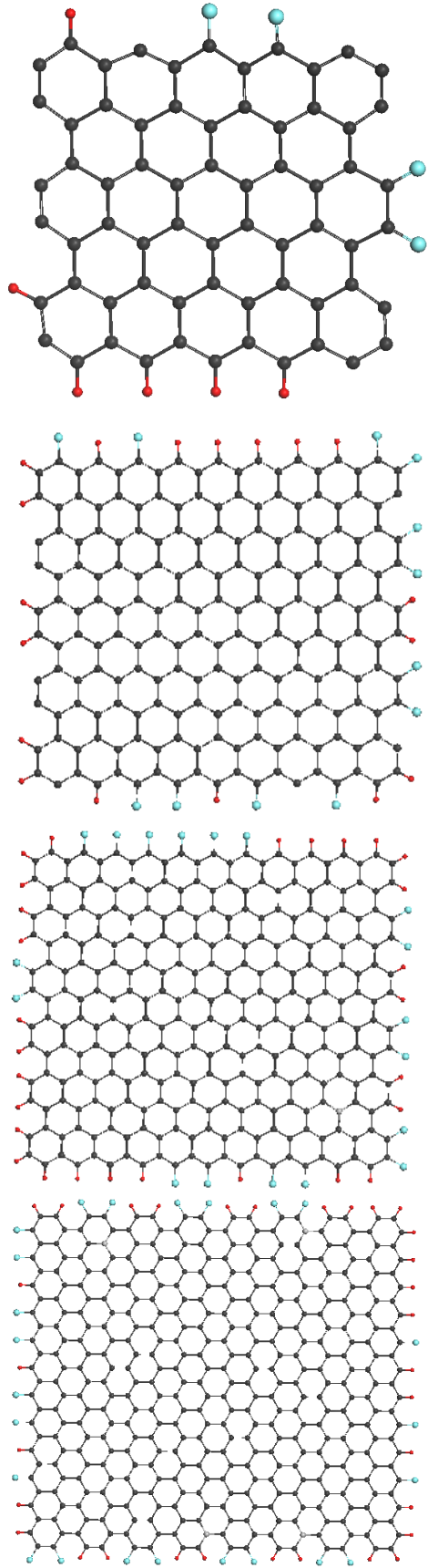

**Figure 4.** BSU models for sequential graphitization of amorphous carbon like shungite carbon. Graphene domains top down: (5,5)NGr (nanographene), (9,9)NGr, (11,11)NGr, (15,12)NGr of lateral dimensions 1.2 nm, 2.2 nm, 2.7 nm, and 3.3 nm, respectively. Dark gray, red and blue balls depict carbon, hydrogen, and oxygen atoms. The chemical composition of the species is described

by an average per one benzenoid formula $C_6O_{0.36}H_{0.55}$ in accordance with chemical content determined in [2].

There might be various reasons for fragmentation, including mechanical impact, chemical reaction, temperature shock, exposure to hard radiation, etc., which can be easily traced by the history of the production of the ACs samples. The fragments ensure the short-range order of the solids and are characterized by large variety with respect to not only different classes of ACs provided by different history and/or technology of their production, but the same class as well. The variety concerns the BSUs size, shape, variation of chemical content, and, which is the most important, the distribution of heteroatoms in the BSU framing area (chemical necklace) at fixed atomic percentage in average. Thus, models presented in Fig. 4, are only 'one snapshot' of communities related to possible permutations of hydrogen and oxygen atoms in the framing area that have no number.

Another distinctive feature of *sp²* ACs lies in the radical nature of their BSU fragments. As shown [2, 3], BSUs of ~ 2nm in size are molecular radicals whose chemical activity is concentrated at non-terminated edge carbon atoms (these atoms are clearly seen on the top of Fig. 2) [2]. However, as can be seen from the figure, increasing the molecule size leads to decreasing edge atoms number due to which, at a fixed atomic composition C: O: H, the number of non-terminated atoms and the degree of the associated radicalization decrease. The weakening of the radicalization of BSU as its size grows, in particular, explains the well-known empirical fact that enforced additional fragmentation leading to nanostructuring of initial synthetic *sp²* ACs significantly enhances the yield of the reaction when these amorphics are used as carbocatalysts [44, 45].

**IV. CONCLUSION**

Structural and analytical studies of *sp²* amorphous carbons were analyzed from the general concepts of the solid-state amorphicity, due to which they can be attributed to molecular amorphics of a new type. The solids are not characterized by structurally and atomically standard molecules but their basic structure units present graphene domains framed with heteroatom necklaces. The domains are resulted from the graphite and/or graphene fragmentation, which itself becomes a distinguished feature of the *sp²* carbon amorphization, which can be called *enforced fragmentation* of honeycomb canvas. Chemical reactions occurred at the graphene domain edges are suggested to be one of the most important causes of the fragmentation stabilization. Thus originated fragments, becoming the basic structural units (BSUs) of AC, are of particular kind presenting size-restricted graphene domains in the halo of heteroatom necklaces. The weak wdW interaction between the BUSs makes them the main defendants for the numerous properties of the *sp²* amorphous carbon solids.

**Acknowledgements**

The study was performed as a part of research topics of the Institute of Geology of Komi Science Center of the Ural Branch of RAS (No. AAAA-A17-117121270036-7). The publication has been prepared with the support of the "RUDN University Program 5-100".